\def\v_op{ \hat{\mathbf v} }
\def\sigmabar{ \bar{\sigma} }

\documentclass[aps,prl,twocolumn,groupedaddress,showpacs]{revtex4}

\usepackage{graphicx}% Include figure files
\usepackage{dcolumn}% Align table columns on decimal point
\usepackage{bm}% bold math

\begin{document}

\title{Quantum Monte Carlo method using phase-free random walks\\ 
with Slater determinants}

\author
{Shiwei Zhang and Henry Krakauer}

\affiliation
{Department of Physics, College of William and Mary,
Williamsburg, VA 23187-8795}

\date{\today}

\begin{abstract}
We develop a quantum Monte Carlo method for many fermions that allows
the use of {\em any\/} one-particle basis. It projects out the ground
state by random walks in the space of Slater determinants.
An approximate approach is formulated to control the phase problem
with a trial wave function $|\Psi_T\rangle$. Using plane-wave basis
and non-local pseudopotentials, we apply the method to Si atom, dimer,
and 2, 16, 54 atom (216 electrons) bulk supercells.  Single Slater
determinant wave functions from density functional theory calculations
were used as $|\Psi_T\rangle$ with no additional optimization.  The
calculated binding energy of Si$_2$ and cohesive energy of bulk Si are
in excellent agreement with experiments and are comparable to the best
existing theoretical results.

\end{abstract}


\maketitle

%\begin{multicols}{2}
%\narrowtext

Quantum Monte Carlo (QMC) methods based on auxiliary fields (AF) are
used in areas spanning condensed matter physics, nuclear physics, and
quantum chemistry. These methods \cite{BSS,Koonin} allow essentially
exact calculations of ground-state and finite-temperature equilibrium
properties of interacting many fermion systems. The required CPU time
scales in principle as a power law with system size, and the methods
have been applied to study a variety of problems including the Hubbard
model, nuclear shell models, and molecular electronic structure.  The
central idea of these methods is to write the imaginary-time
propagator of a many-body system with two-body interactions in terms
of propagators for independent particles interacting with external
auxiliary fields. The independent particle problems are solved for
configurations of the AF and averaging over different AF
configurations is then performed by Monte Carlo (MC) techniques.

QMC methods with auxiliary fields have several appealing features.
For example, they allow one to choose {\em any\/} one-particle basis
suitable for the problem, and to fully take advantage of
well-established techniques to treat independent particles. Given the
remarkable development and success of the latter \cite{DFT}, 
it is clearly very
desirable to have a QMC method that can use exactly the same machinery
and systematically include correlation effects by simply building
stochastic ensembles of the independent particle solutions. Vigorous
attempts have been made from several fields to explore this
possibility 
\cite{Silvestrelli93,shiftcont_rom97,Wilson95,AFQMC_appl}.

A significant hurdle exists, however: except for special cases (e.g.,
Hubbard), the two-body interactions will require auxiliary fields that
are {\em complex\/}. As a result, the single-particle orbitals become
complex, and the MC averaging over AF configurations becomes an
integration over complex variables in many dimensions. A phase problem
thus occurs which ultimately defeats the algebraic scaling of MC and
makes the method scale exponentially. This is analogous to but more
severe than the fermion sign problem with real AF \cite{loh90,Zhang}
or in real-space methods \cite{schmidt84}.  No satisfactory, general
approach exists to control the phase problem. As a result, only small
systems or special forms of interactions can be treated.

In this paper we address this problem. We develop a method for
many-fermions that allows the use of any one-particle basis. It
projects out the ground state by random walks in the space of Slater
determinants. The phase problem is eliminated with an approximation
that relies on a trial wave function $|\Psi_T\rangle$.  We demonstrate
the method by applying it to electronic systems using a plane-wave
basis and non-local pseudopotentials, which can be implemented
straightforwardly in this method.  We calculate the binding energy of
Si$_2$ and the cohesive energy of bulk Si using fcc supercells
consisting of up to 54 atoms (216 electrons).  These calculations
represent the first application of AF-based QMC to solids.  The
results are in excellent agreement with experiments and are comparable
to the best existing theoretical results. Particularly worth noting is
that our results were obtained with a trial wave function which is a
single Slater determinant formed by orbitals from density functional
theory (DFT) calculations (with the local density approximation
(LDA)), with no additional parameters or optimization.

The Hamiltonian for any many-fermion system with two-body interactions
can be written in any one-particle basis in the general form
\begin{equation}
{\hat H} ={\hat H_1} + {\hat H_2}
= \sum_{i,j}^N {T_{ij} c_i^\dagger c_j}
   + {1 \over 2} 
\sum_{i,j,k,l}^N {V_{ijkl} c_i^\dagger c_j^\dagger c_k c_l},
\label{eq:H}
\end{equation}
where $N$ is the size of the chosen one-particle basis, and
$c_i^\dagger$ and $c_i$ are the corresponding creation and
annihilation operators.  Both the one-body ($T_{ij}$) and two-body
matrix elements ($V_{ijkl}$) are known.

To obtain the ground state $\left| \Psi_G \right\rangle$ of ${\hat
H}$, QMC methods use the imaginary time propagator $e^{-\tau {\hat
H}}$ acting on a trial wave function $\left| \Psi_T \right\rangle$:
$\lim_{n
\to \infty} (e^{-\tau {\hat H}})^n \left| \Psi_T \right\rangle \propto
\left| \Psi_G \right\rangle$.  $\left| \Psi_T \right\rangle$ must not
be orthogonal to $\left| \Psi_G \right\rangle$, and we will assume
that it is of the form of a single Slater determinant or a linear
combination of Slater determinants.  The time step $\tau$ is chosen to
be small enough so that ${\hat H_1}$ and ${\hat H_2}$ in the
propagator can be accurately separated with the Trotter 
decomposition.

The propagator $e^{-\tau {\hat H_1}}$ is the exponential of a one-body
operator. A propagator of this form acting on a Slater determinant is
straightforward to calculate, and it simply yields another
determinant. The two-body propagator $e^{-\tau {\hat H_2}}$ can be
expressed as an integral of propagators of this form, as follows.  Any
two-body operator can be written as a quadratic form of one-body
operators:
${\hat H_2} = - {1\over 2}\sum_\alpha \lambda_\alpha
{\hat v_\alpha}^2$, 
where $\lambda_\alpha$ is a real number and
${\hat v_\alpha}$ is a one-body operator.
The Hubbard-Stratonovich (HS) transformation \cite{HS} 
then allows us to write
\begin{equation}
   e^{-\tau{\hat H_2}}
= \prod_\alpha \Bigg({1\over \sqrt{2\pi}}\int_{-\infty}^\infty
            e^{-\frac{1}{2} \sigma_\alpha^2}\:
           e^{\sqrt{\tau}\,\sigma_\alpha\,
\sqrt{\lambda_\alpha}\,{\hat v_\alpha}}\:d\sigma_\alpha\Bigg).
\label{eq:HStrans}
\end{equation}
Introducing vector representations $\sigma\equiv \{\sigma_1,\sigma_2,
\cdots\}$ and $\v_op=\{ \sqrt{\lambda_1}\,{\hat v_1},
\sqrt{\lambda_2}\,{\hat v_2}, \cdots\}$, we have the 
desired form 
\begin{equation}
e^{-\tau{\hat H}}
=\int P(\sigma)\:B(\sigma)\:d\sigma,
\label{eq:prop}
\end{equation}
where $P(\sigma)$ is the normal distribution in Eq.~(\ref{eq:HStrans})
and 
\begin{equation}
B(\sigma)\equiv
e^{-\tau {\hat H_1}/2}
\:e^{\sqrt{\tau} \sigma\cdot {\hat{\mathbf v}}}
\:e^{-\tau {\hat H_1}/2}
\label{eq:Bdef}
\end{equation}
is a one-body propagator.

The imaginary-time propagation thus requires evaluating the
multidimensional integral in Eq.~(\ref{eq:prop}) over time slices $n$
and the corresponding auxiliary fields.  MC techniques are the only
way to evaluate such integrals efficiently.  We use a random walk
approach \cite{Zhang}.  In each step, a walker $|\phi\rangle$, which
is a single Slater determinant, is propagated to a new position $
|\phi^\prime\rangle$: $ |\phi^\prime(\sigma)\rangle=
B(\sigma)|\phi\rangle$,
where $\sigma$ is a random variable sampled from $P(\sigma)$.  After a
sufficient number of steps (iterations), the ensemble of random
walkers is a MC representation of the ground-state wave function:
$|\Psi_G\rangle \doteq \sum_{\phi^\prime} |\phi^\prime\rangle$.

\begin{figure}
\includegraphics[width=0.42\textwidth]{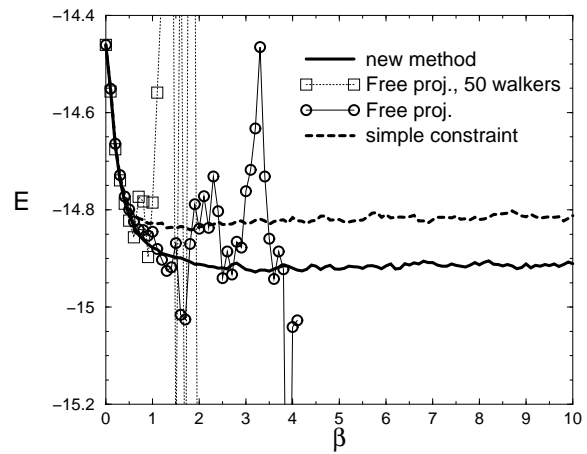}
\caption{Illustration of the phase problem and constraints to 
control it. The total valence energy (in Ry) of an fcc Si primitive
cell (2 atoms) is shown as a function of projection time
$\beta=n\tau$, with $\tau=0.05$ Ry$^{-1}$.  Unless otherwise
indicated, $10,000$ walkers are used.  Increasing the number of
walkers from $50$ to $10,000$ only slightly delays the onset of the
phase problem. Simple generalization of the constraint that worked
well for real determinants leads to poor results. The new method gives
accurate results (note the agreement with the solid free projection
curve, which is exact, until the latter becomes too noisy at $\beta
\sim 1.5$). }
\label{fig1}
\end{figure}

In general $\lambda_\alpha$ cannot be made all positive in
Eq.~(\ref{eq:HStrans}) \cite{note_shift_potential}. The one-body
operators $\v_op$ are therefore complex.  As the projection proceeds,
the orbitals in the random walkers will become complex. As a result,
the statistical fluctuations in the MC representation of
$|\Psi_G\rangle$ increase exponentially with projection time
$\beta\equiv n\tau$.  This is the phase problem referred to
earlier. It is of the same origin as the sign problem that occurs when
$B(\sigma)$ is real. The phase problem is more severe, however,
because for each $|\phi\rangle$, instead of a $+|\phi\rangle$ and
$-|\phi\rangle$ symmetry \cite{Zhang}, there is now an infinite set
$\{ e^{i\theta} |\phi\rangle\}$ ($\theta \in [0,2\pi)$) from which the
random walk cannot distinguish. At large $\beta$, the phase of each
$|\phi\rangle$ becomes random, and the MC representation of
$|\Psi_G\rangle$ becomes dominated by noise. This problem is generic,
and the same analysis would apply if we had chosen, instead of the
random walk, the standard AF QMC sampling approach \cite{Koonin}.  In
Fig.~\ref{fig1}, the curves labeled ``free projection'' illustrate the
phase problem.

Existing fixed-node type approximations have often worked very well to
control the sign/phase problem in real space \cite{Foulkes01,Ortiz93}
or in Slater determinant space when the propagator is real
\cite{Zhang}.  The phase problem here is unique because not only do
the determinants acquire overall phases, but the internal structures
of their orbitals become complex.  The real-space analogy would be to
have walkers whose coordinates become complex.  This makes
straightforward generalization of existing approaches ineffective. For
example, similar to the constrained path approximation \cite{Zhang} we
could impose the condition ${\rm Re}\langle\Psi_T |
\phi\rangle >0 $. 
Or, in the spirit of the fixed-phase approximation in real 
space \cite{Ortiz93} we
could project the walker by including a factor ${\rm cos}(\Delta\theta)$
in the weight,
where $\Delta\theta$ is the phase of $\langle\Psi_T
|\phi^\prime\rangle/\langle\Psi_T |\phi\rangle$.  They give similar
results and do not work well \cite{Wilson95}. 
The former is shown in Fig.~\ref{fig1}
(``simple constraint''). Importance sampling with ${\rm Re}\langle\Psi_T |
\phi\rangle$ or $|\langle\Psi_T |\phi\rangle|$ does not change the 
results \cite{ZKunpublished}.

To formulate a new method that can better separate the overall phase
from the determinant, we first borrow from the idea of importance
sampling \cite{Kalos74}, although our choice of the so-called
importance function, $\langle \Psi_T|\phi\rangle$, is actually {\em
complex\/}.  We modify Eq.~(\ref{eq:prop}) to obtain the following new
propagator for $|\phi\rangle$:
\begin{equation}
\int \langle \Psi_T|\phi^\prime(\sigma-\sigmabar)\rangle\:
P(\sigma-\sigmabar) B(\sigma-\sigmabar) 
{1 \over \langle\Psi_T | \phi\rangle}\:d\sigma,
\label{eq:imp_sigmabar}
\end{equation}
where we have included a constant shift \cite{shiftcont_rom97}
$\sigmabar$ in the integral in Eq.~(\ref{eq:prop}), which does not
affect the equality.  Eq.~(\ref{eq:imp_sigmabar}) can be re-written as
\begin{equation}
\int P(\sigma)\:W(\sigma,\phi)\:B(\sigma-\sigmabar)\:d\sigma,
\label{eq:prop_imp}
\end{equation}
where 
\begin{equation}
W(\sigma,\phi)\equiv {\langle \Psi_T|\phi^\prime(\sigma-\sigmabar)\rangle
\over \langle\Psi_T | \phi\rangle}
e^{\sigma \cdot \sigmabar - {\sigmabar\cdot \sigmabar \over 2}}.
\label{eq:wt_raw}
\end{equation}

The new propagator in Eq.~(\ref{eq:prop_imp}) defines a new random
walk. In each step the walker $|\phi\rangle$ is propagated to
$|\phi^\prime\rangle$ by $B(\sigma-\sigmabar)$:
$|\phi^\prime(\sigma-\sigmabar)\rangle=B(\sigma-\sigmabar) |
\phi\rangle$, where $\sigma$ is again sampled from $P(\sigma)$.
$W(\sigma,\phi)$ is a c-number which can be accounted for by having
every walker carry an overall weight factor and updating them
according to: $w_{\phi^\prime}=W(\sigma,\phi) w_\phi$.  Formally the
MC representation of $|\Psi_G\rangle$ in the new random walk is:
\begin{equation}
|\Psi_G\rangle \doteq \sum_{\phi^\prime} w_{\phi^\prime}
{|\phi^\prime\rangle \over \langle\Psi_T | \phi^\prime\rangle}.
\label{eq:wf_MC_imp}
\end{equation}

For any choice of the shift $\sigmabar$, the new random walk is an
exact procedure to realize the imaginary time propagation, in the
sense of Eq.~(\ref{eq:wf_MC_imp}). The optimal choice of $\sigmabar$
is determined by minimizing the fluctuation of $W(\sigma,\phi)$ with
respect to $\sigma$. To ${\cal O}(\sqrt\tau)$ this yields
\begin{equation}
\sigmabar =
- \sqrt{\tau}
{\langle\Psi_T|\v_op|\phi\rangle \over \langle\Psi_T | \phi\rangle}.
\label{eq:FB}
\end{equation}
With this choice the leading $\sigma$-dependent term in $W$ is reduced
to ${\cal O}(\tau)$ and, by expanding $B(\sigma-\sigmabar)$ in
$|\phi^\prime\rangle$ in Eq.~(\ref{eq:wt_raw}), we can manipulate $W$
into the following form:
\begin{equation}
W(\sigma,\phi) \doteq 
\exp\bigg[-\tau  
{\langle\Psi_T|\hat{H}|\phi\rangle \over \langle\Psi_T | \phi\rangle}\bigg]
\equiv \exp[-\tau E_L(\phi)],
\label{eq:El}
\end{equation}
where the term $E_L$ parallels the local energy in real-space QMC
methods. Both $E_L$ and the shift $\sigmabar$ in Eq.~(\ref{eq:FB}) are
independent of any overall phase factor of $|\phi\rangle$.

The weight of the walker in the new random walk is determined by
$E_L$.  In the limit of an exact $|\Psi_T\rangle$, $E_L$ is a real
constant, and the weight of each walker remains real.  The so-called
mixed estimate for the energy is phaseless:
\begin{equation}
E_G = 
{\langle\Psi_T|\hat{H}|\Psi_G\rangle \over \langle\Psi_T | \Psi_G\rangle}
\doteq
{\sum_{\phi^\prime} w_{\phi^\prime} E_L({\phi^\prime}) 
\over 
\sum_{\phi^\prime} w_{\phi^\prime}}.
\label{eq:mixed_w_EL}
\end{equation}
With a general $|\Psi_T\rangle$ which is not exact, a natural
approximation is to replace $E_L$ in Eq.'s (\ref{eq:El}) and
(\ref{eq:mixed_w_EL}) by its real part, ${\rm Re} E_L$.  We have thus
obtained a phaseless formalism for the random walk, with real and
positive weights in Eq.'s~(\ref{eq:wf_MC_imp}) and
(\ref{eq:mixed_w_EL}).

Despite this, an additional constraint is still required.
To illustrate the problem we consider the overlap
$\langle\Psi_T|\phi^\prime\rangle$ during the random walk. Let us
denote the phase of $\langle\Psi_T
|\phi^\prime(\sigma-\sigmabar)\rangle/\langle\Psi_T |\phi\rangle$ by
$\Delta\theta$, which is in general non-zero (of order $ -\sigma {\rm
Im}\sigmabar$).  This means that, the walkers will undergo a random
walk in the complex plane defined by
$\langle\Psi_T|\phi^\prime\rangle$.  At large $\beta$ they will
therefore populate the complex plane symmetrically, independent of
their initial positions.  It is useful to contrast the situation with
the special case of a {\em real\/} $\v_op$.  For any $\v_op$ the shift
$\sigmabar$ diverges as a walker approaches the origin in the complex
plane, i.e., as $\langle\Psi_T|\phi^\prime\rangle
\rightarrow 0$. The effect of the divergence is to move the walker
away from the origin. With a {\em real\/} $\v_op$, $\Delta\theta=0$
and the random walkers move only on the real axis. If they are
initialized to have positive overlaps with $|\Psi_T\rangle$,
$\sigmabar$ will ensure that the overlaps remain positive throughout
the random walk, much like in fixed-node diffusion Monte Carlo (DMC)
in real space. Thus in this case the phaseless formalism reduces to
the constrained path Monte Carlo method of Ref.~\cite{Zhang}, and it
alone is sufficient to control the sign problem. For a {\em complex\/}
$\v_op$, however, the random walk is ``rotationally invariant'' in the
complex plane, and the divergence of $\sigmabar$ is not enough to
prevent the build-up of a finite density at the origin. Near the
origin the local energy $E_L$ diverges, which causes diverging
fluctuations in the weights of walkers.  To address this we make an
additional approximation. We project the random walk to
``one-dimension'' and multiply the weight of each walker in each step
by $\max\{0,\cos(\Delta\theta)\}$.
Imposing instead ${\rm Re}\langle\Psi_T|\phi^\prime\rangle> 0$ 
gave similar results, but with somewhat larger variance.

We apply the new method to Si atom, molecule, and bulk.  The Si$^{4+}$
ions are represented by a norm-conserving LDA Kleinman-Bylander (KB)
non-local pseudopotential \cite{rappe}.  We use periodic boundary
conditions, and a plane-wave basis with a kinetic energy cut-off
$E_{\rm cut}=12.25$ Ry.  The error resulting from $E_{\rm cut}$ was
estimated through LDA calculations and is smaller than the MC
statistical errors.  The pseudopotential can be applied in essentially
the same way as in plane-wave-based LDA calculations
\cite{ZKunpublished}. Calculations involving $\v_op$ and the local
part of the pseudopotential are efficiently handled using fast Fourier
transforms. The separable KB form of the non-local pseudopotential
makes its application as efficient as in LDA plane-wave codes.  Our
$|\Psi_T\rangle$ is a single Slater determinant consisting of LDA
orbitals.

In Table \ref{tab1}, we show results for the atom and molecule. 
Additional calculations with $a=22a_B$
supercells show that finite-size errors at $a=19 a_B$ were smaller
than the MC statistical errors.
Our calculated Si$_2$ binding energy is in excellent agreement with
the experimental value \cite{Schmude}.

\begin{table}
\caption{Total valence energies of Si and Si$_2$, and binding energy of
Si$_2$. 
The Si$_2$ ground state is $^3\Sigma_g^-$ (electronic
configuration $5\uparrow\,3\downarrow$). Calculations were 
done at the experimental
equilibrium bond length of $4.244 a_B$, in a cubic supercell with
$a=19a_B$ (4945 plane waves). Energies are in
eV. Error bars are in the last digit and are 
in parentheses. 
}
\label{tab1}
\begin{ruledtabular}
\begin{tabular}{lccc}
           &  Si               &  Si$_2$         &  Si$_2$ $E_B$  \\  \hline
LDA        & $ -102.648\ \ $   & $-209.175\ \ $  &  $3.879\ \ \ $   \\  
QMC        & $ -103.45(2)$     & $-210.03(7)$    &  $3.12(8)\ $     \\
Experiment &                   &                 &  $3.21(13)$
\end{tabular}
\end{ruledtabular}

\end{table}

In the bulk calculations, we use fcc supercells consisting of 2, 16,
54 atoms (5209 plane waves).  As Fig.~\ref{fig1} shows, the new method
leads to a large improvement.  Results for 16 and 54 atoms are shown
in Table \ref{tab2}. Our calculation for 54 atoms took several days on
20 Compaq Alpha 667 MHz processors. For the bulk cohesive energy, we
first included a correction for the independent-particle finite-size
error from the LDA results. We then corrected for the remaining
Couloumb finite-size error \cite{note_CFSE} using the results of Kent
{\it et~al.\/} \cite{Kent99}.
Our result is
again in excellent agreement with the 
experimental value (from Ref.~\cite{Foulkes01}). It also
compares very well with the result of 
a recent fixed-node DMC
calculation \cite{leung99}, which also used a 54-atom supercell and
gave $4.63(2)$ eV per atom after similar finite-size 
and zero-point energy corrections.

\begin{table}
\caption{Cohesive energy 
of bulk Si. 
Calculations are done for fcc supercells with 16 and 54 atoms, at
$a_{\rm exp}=5.43$\AA.  QMC result at $\infty$ is from 54 atoms and
includes two finite-size corrections: (i) an independent-particle
correction of $0.311$ eV from LDA and (ii) an additional Couloumb
correction of $-0.174$ eV from Ref.~\cite{Kent99, Kent_Priv}.  A
zero-point energy correction of $-0.061$ eV was also added to the
calculated results at $\infty$.  Energies are in eV per atom. Error
bars are in the last digit and are in parentheses. }
\label{tab2}
\begin{ruledtabular}
\begin{tabular}{lccc}
           &  16               &  54         &   $\infty$  \\  \hline
LDA        & $3.836\ \ $  & $4.836\ \ $  &  $5.086\ \ $   \\
QMC        & $3.79(4)$    & $4.51(3)$    &  $4.59(3)$     \\
Experiment &              &              &  $4.62(8)$
\end{tabular}
\end{ruledtabular}

\end{table}

Without an exact solution to the sign/phase problem, reducing the
reliance on trial wave functions is clearly of key importance to
increasing the predictive power of QMC.  For continuum electronic
systems such as our test cases above, fixed-node DMC has often been
the most accurate theoretical method \cite{Foulkes01}.  It is
encouraging that the new method, using simple LDA trial wave
functions, gave comparable
results to DMC. For similar supercells DMC often uses trial wave
functions with 30-100 additional parameters \cite{Foulkes01}.
Obtaining a good enough $|\Psi_T\rangle$ is instrumental to a
successful DMC calculation, and often constitutes a substantial
effort.  The quality of $|\Psi_T\rangle$ controls the systematic
errors from the fixed-node approximation and the variance.  It also
affects errors due to the locality approximation \cite{Mitas91}, which
has been employed by most DMC calculations with non-local
pseudopotentials.  In the new method the latter approximation is
eliminated. It remains to be seen whether the present method
could lead to more accurate results than
fixed-node DMC for continuum systems. 
This has been possible in some cases \cite{Zhang99}
with real AF in the Hubbard model.

We have presented a general framework. Various possibilities exist for
further improvement of the method.  An improved $|\Psi_T\rangle$ will
give improved results. The freedom to choose the one-particle basis
and the form of HS transformation, both of which can impact the
quality of the results,
offers significant opportunities.  For periodic systems it should be
possible to generalize the formalism to allow ${\mathbf k}$-point
sampling. 

In conclusion, we have described a method for ground-state QMC
calculations that allows the use of any one-particle basis.  The
method is general and applies to any Hamiltonian of the form in
Eq.~(\ref{eq:H}). It provides an approximate way to control the phase
problem in all AF-based QMC methods, while allowing many of their
advantages to be retained that lead to their applications spanning
several areas. We have shown that the method gave accurate results for
systems from an atom to a large supercell, using a simple trial wave
function.

We are very grateful to E.~J.~Walter for help with LDA calculations
and with several programming issues.  We thank P.~Kent for sending us
his DMC data and J.~Carlson for useful conversations.  This work is
supported by the NSF under Grant No.~DMR-9734041 and the Research
Corporation, and by ONR Grant DMR-N000149710049.  We acknowledge
computing support by the Center for Piezoelectrics by Design.

%\bibliography{Jell_L}% Produces the bibliography via BibTeX.

\begin{thebibliography}{10}
\bibitem{note_shift_potential} Shifting the potential can in principle
make $\lambda_\alpha$ positive, but causes large fluctuations (see,
e.g., Ref.~\cite{Silvestrelli93}) and gives poor results when a
constraint is used.
\bibitem{note_CFSE}
  In Ref.~\cite{Kent99}, the DMC calculations used $L$-point sampling,
  but it was pointed out that, after the independent-particle
  correction, the remaining Couloumb finite-size error is
  approximately the same for $\Gamma$-point sampling.
\end{thebibliography}

\end{document}